\documentclass[a4paper, amsfonts, amssymb, amsmath, reprint, showkeys, nofootinbib, twoside]{revtex4-1}
\usepackage[english]{babel}
\usepackage[utf8]{inputenc}
\usepackage[colorinlistoftodos, color=green!40, prependcaption]{todonotes}
\usepackage{amsthm}
\usepackage{ulem}
\usepackage{multirow} 
\usepackage{xcolor}
\usepackage{graphicx}
\usepackage{caption}
\usepackage[T1]{fontenc}
\usepackage[pdftex, pdftitle={Article}, pdfauthor={Author}]{hyperref} % For hyperlinks in the PDF

\definecolor{purple}{rgb}{0.58,0.0,0.83}

\definecolor{blue(pigment)}{rgb}{0.2, 0.2, 0.6}

\usepackage{scalerel}
\usepackage{tikz}
\usetikzlibrary{svg.path}

\definecolor{orcidlogocol}{HTML}{A6CE39}
\tikzset{
  orcidlogo/.pic={
    \fill[orcidlogocol] svg{M256,128c0,70.7-57.3,128-128,128C57.3,256,0,198.7,0,128C0,57.3,57.3,0,128,0C198.7,0,256,57.3,256,128z};
    \fill[white] svg{M86.3,186.2H70.9V79.1h15.4v48.4V186.2z}
                 svg{M108.9,79.1h41.6c39.6,0,57,28.3,57,53.6c0,27.5-21.5,53.6-56.8,53.6h-41.8V79.1z M124.3,172.4h24.5c34.9,0,42.9-26.5,42.9-39.7c0-21.5-13.7-39.7-43.7-39.7h-23.7V172.4z}
                 svg{M88.7,56.8c0,5.5-4.5,10.1-10.1,10.1c-5.6,0-10.1-4.6-10.1-10.1c0-5.6,4.5-10.1,10.1-10.1C84.2,46.7,88.7,51.3,88.7,56.8z};
  }
}

\newcommand\orcidicon[1]{\href{https://orcid.org/#1}{\mbox{\scalerel*{
\begin{tikzpicture}[yscale=-1,transform shape]
\pic{orcidlogo};
\end{tikzpicture}
}{|}}}}

\begin{document}

\title{Exploring the cosmological degeneracy between decaying dark matter model and viscous $\Lambda$CDM}

\author{Gilberto Aguilar-P\'erez$^1$\orcidicon{0000-0001-6821-4564}}
\email{gilaguilar@uv.mx}

\author{Ana A. Avilez-López$^{2,3}$\orcidicon{0000-0003-2223-4716}}
\email{ana.avilezlopez@correo.buap.mx}

\author{Miguel Cruz$^1$\orcidicon{0000-0003-3826-1321}}
\email{miguelcruz02@uv.mx}

\affiliation{$^1$Facultad de F\'{\i}sica, Universidad Veracruzana 91097, Xalapa, Veracruz, M\'exico,\\
$^2$Facultad de Ciencias F\'{\i}sico Matem\'aticas, Benem\'erita Universidad Aut\'onoma de Puebla, Apdo. Postal 1152, Puebla, Pue., M\'exico, \\
$^3$Centro Internacional de F\'isica Fundamental, Benem\'erita Universidad Aut\'onoma de Puebla, Apdo. Postal 1152, Puebla, Pue., M\'exico.}

\date{\today}

\begin{abstract}
In the context of a homogeneous and isotropic universe, we consider the degeneracy condition at the background level between two scenarios in which processes out of equilibrium are possible; this consideration allows us to deviate from the perfect fluid description, and in this case bulk viscosity represents a viable candidate to describe entirely such effects. The cosmological model describing an unstable dark matter sector is mapped into a slight modification of the $\Lambda$CDM model characterized by a viscous dark matter sector; under this consideration our description does not depend on a specific formulation of viscous effects and these can be fully reconstructed and characterized by the parameter that determines the decay ratio of dark matter. However, in this new scenario the cosmic expansion is influenced by the viscous pressure and the dark energy sector given by the cosmological constant is translated into a dynamical one. As a consequence of our formulation, the test against observations of the model indicates consistency with quintessence dark energy, but the crossing of the phantom divide can be accessible.       
\end{abstract}

%\pacs{98.80.Cq}

\maketitle

%%%%%%%%%%%%%%%%%%%%%%%%%%%%%%%%%%
\section{Introduction} 
%%%%%%%%%%%%%%%%%%%%%%%%%%%%%%%%%%%

It is well known that the cosmological concordance model, or simply $\Lambda$CDM, is based on the existence of a non-relativistic fluid known as cold dark matter and a cosmological constant (CC) which is interpreted as vacuum energy and represents the dark energy sector. In this scenario dark matter dominates the growth of structures and deceleration at early stages of cosmological evolution, while the CC drives the current accelerated expansion. The amount of dark matter at the recombination epoch has been inferred from the Cosmic Microwave Background (CMB) observations \cite{planck1518} and its comoving density is constant throughout the cosmic evolution together with the equation of state $\omega_{m}=0$; besides, the linearized density and velocity perturbations satisfy the continuity and pressureless Euler equations. However, as attractive as the simplicity of the $\Lambda$CDM model may be, the high-definition astrophysical data indicate that we require to look beyond this model in order to describe the observable universe more accurately \cite{riess2024}.\\

Regarding the aforementioned necessity and leaving the dark energy sector aside for the moment, the most simple assumption to explore is the consideration of deviations from the usual description mentioned above for dark matter. As first example we can mention the generalized dark matter model given by Hu \cite{hu}, in which $\omega_{m} \neq 0$ through the cosmic history, this proposal was extensively studied and extended to the context of interacting dark sector in \cite{hu2, hu3}. The results reported in Ref. \cite{hu3} indicated that the dark matter abundance is strictly positive and $\omega_{m}$ is consistent with zero. But an interesting characteristic of this model is that the non-linearities at small scales can alter the cosmological background and large scale linear perturbations, these effects are characterized by the creation of an effective pressure and viscosity in the fluid, the physical interpretation is that viscosity will act to damp out the velocity fluctuations. The emergence of viscosity (and effective pressure) in an initially pressureless perfect fluid was also confirmed in the context of an effective field theory of large-scale structure \cite{large}, i.e., the long-wavelength universe behaves as a viscous fluid coupled to gravity.\\

Another interesting possibility to consider is the relaxation of the comovingly constant dark matter density condition. A captivating approach is the decay of a fraction of dark matter into relativistic particles of the dark sector, i.e., dark radiation. This is the most simple scenario and seems to resolve several discrepancies of standard cold dark matter; see, for instance, \cite{takahashi}. On the other hand, such a conversion has been the subject of exhaustive investigation since alleviates some of the tensions that plague the $\Lambda$CDM model; see Refs. \cite{lesgourgues, das1, das2, bringmann, vattis} and references therein. Additionally, the conversion of dark matter into dark radiation can also be found in the merging of primordial black holes emitting gravitational waves \cite{bh}, this can promote the use of the recent results released by the LIGO Collaboration \cite{ligo} in order to be implemented in this kind of cosmological scenario. The nature of dark radiation has not been fully understood so far. However, some studies allow dark radiation to be comprised by sterile neutrinos as well as active neutrinos with a non-thermal distribution, in such a case neutrinos are expected to behave as relativistic particles with effective sound speed and viscosity parameter, see Ref. \cite{lesgourgues2}; the existence of bulk viscosity in the universe was demonstrated long ago using a gas of neutrinos \cite{neutribulk}. Notice that in both scenarios mentioned here for modified dark matter, the emergence of a viscous contribution in the cosmic fluid is a common factor.\\ 

In this sense, in order to study the cosmological viability of both approaches to describe the current accelerated expansion; in this work we consider a mapping at the background level between a specific cosmological model in which dissipative processes are present due to the slow decay of dark matter into dark radiation with a dark energy sector characterized by a constant parameter state, $\omega_{\mathrm{de}}$ \cite{wcdm} and a slightly modified version of the $\Lambda$CDM model, in this case the modification is given by considering a viscous dark matter sector; due to the contribution of the viscous pressure, $\Pi$, the condition $\omega_{m} \neq 0$ is covered. As a result of this degeneracy condition, the viscous effects considered in the $\Lambda$CDM model can be characterized by a single parameter, $\alpha$, which in turn determines the decay ratio for dark matter and at the same time the CC becomes dynamical. This latter situation can be useful to circumvent the well-known {\it CC problem}: the great discrepancy found between the values of $\Lambda$ (interpreted as vacuum energy) obtained from quantum field theory and observations indicates that the nature of dark energy can not be entirely comprised by the CC \cite{lambda}. The conversion from the constant behavior to a dynamical one for the dark energy sector is obtained naturally in this formulation, but is the most simple phenomenological way to extend and explain the nature of dark energy and has been widely explored in the literature by considering parameterizations of the form $\Lambda=\Lambda(a)$ and/or $\Lambda=\Lambda(H)$, being $a$ the scale factor and $H$ the Hubble parameter; see, for instance Refs. given in \cite{vcc}. Additionally, this procedure allows one to reconstruct the form of the bulk viscosity without the need of solving the constitutive equations that characterize the viscous effects, in general, these equations are hard to solve in this kind of scenario \cite{is, nonlinear}. The presence of the term $\Pi$ reduces the total pressure of the fluid and, consequently, contributes to the accelerated cosmic expansion.\\

We highlight the fact that both models to be considered in this work have in common that their processes occur out of equilibrium, and the mapping can be done given the degeneracy of the expansion of the universe at background level following the methodology given in Ref. \cite{velten} for an homogeneous and isotropic universe.\\ 

The structure of the article is as follows: in the next Section the generalities of the effective cosmological models to be used are described, as well as the properties of the thermodynamics of irreversible processes involved, we comment the similarities at background dynamics level between bulk viscosity and the matter creation scenario. We specify the main properties of the decaying dark matter model and the modified $\Lambda$CDM model. In Section \ref{sec:data} a Bayesian analysis is performed in order to test the model against the observational data. In this analysis, the parameters of the model are constrained and their uncertainties are estimated. Finally in section \ref{sec:final} we give some final comments for our work. 

%%%%%%%%%%%%%%%%%%%%%%%%%%%%%%%%%%%%%%%%%%%%%%%%%%%%%%%%%%%
\section{Cosmological models}
\label{sec:cm}
%%%%%%%%%%%%%%%%%%%%%%%%%%%%%%%%%%%%%%%%%%%%%%%%%%%%%%%%%%

This section is devoted to describing the general aspects of the cosmological models to consider in our analysis for a Friedmann-Lema\^itre-Robertson-Walker (FLRW) spacetime with null spatial curvature; $8\pi G = c = 1$ units will be used.

\subsection{Some aspects of irreversible thermodynamics}

In general, the evolution of the universe is described by assuming that matter corresponds to a perfect fluid. However, by considering a fluid of this type, some physical features of matter are left out. Such features are relevant if dissipative processes occur. Consequently, under the typical prescription for matter, cosmic evolution turns out to be a reversible process. Within such a standard scenario, physical states of matter are described by local equilibrium variables even when processes for which the non-equilibrium condition is expected, as happens for the expanding fluid case. Clearly, something is missing from its description from the thermodynamic point of view. In order to generalize the description of the cosmic fluid, in this work we consider deviations from local equilibrium variables. A well known example of this kind of schema is to introduce dissipative effects. The evolution of the universe contains several dissipative processes, we mention some: the spontaneous symmetry breaking when gauge bosons acquire mass, the decay of the primordial scalar field into particles, i.e., the reheating stage at the end of inflation, the decoupling of neutrinos from cosmic plasma and the decoupling of photons from matter during the recombination era. Within the astrophysical context the gravitational collapse of local inhomogeneities to form galactic structures is an example in which dissipation is involved. Additionally, the collapse of a radiating star to a neutron star o black hole is another example, since heat flow and viscosity are present. Finally, the accretion of matter around a neutron star or black hole and binary neutron star merger are also examples of dissipative processes; these latter examples provide a good environment to explore the properties of matter at extreme densities and temperatures \cite{maartens, noronha}. \\

For a dissipative fluid, the energy density coincides with the local equilibrium value (this is also valid for other thermodynamics scalars), but the pressure deviates from the local pressure value as follows \cite{maartens}
\begin{equation}
    p_{\mathrm{eff}}=p+\Pi,
    \label{eq:pressure}
\end{equation}
where $p_{\mathrm{eff}}$ is dubbed as the effective non-equilibrium pressure and $\Pi$ denotes the bulk viscous (or non-adiabatic) pressure. The energy balance equation takes the form
\begin{equation}
    \dot{\rho} + 3H(\rho+p_{\mathrm{eff}}) = 0 \ \ \ \longrightarrow \ \ \ \dot{\rho} + 3H(\rho+p+\Pi) = 0,
    \label{eq:energy}
\end{equation}
where the dot denotes derivatives w.r.t. proper time and $H$ is the Hubble parameter. Notice that the above equation can be written as an inhomogeneous equation, being $\Pi$ the source/sink of energy density. Such role for $\Pi$ has an important meaning at thermodynamics level since it also can be though as an entropy catalyst, leading to a non adiabatic cosmological expansion, as we will see below. This latter characteristic represents an interesting cosmological scenario since non-equilibrium phenomena are also present in other cosmological models. For instance, in Ref. \cite{entropy} it was found that the growth of entropy associated to the causal horizon of our universe induces an acceleration that is indistinguishable from the one obtained in the $\Lambda$CDM model for an homogeneous and isotropic cosmology. Based on the aforementioned discussion, our description for the fluid does not satisfy the equilibrium condition. In this case the Friedmann equations are modified in the following form
\begin{align}
& 3H^2 = \rho, \label{eq:fried1}\\
& 2\dot{H} + 3H^2 = -p-\Pi. \label{eq:accel}
\end{align}
It is worthy to mention that the set given by the equations (\ref{eq:energy}), (\ref{eq:fried1}) and (\ref{eq:accel}), resembles the background dynamics associated to more general cosmological scenarios as unimodular gravity. Such scheme allows diffusion processes which induce deviations from the equilibrium condition due to the non-conservation of the energy-momentum tensor, see Ref. \cite{nucamendi} for an interesting review of the model and its role alleviating the $H_{0}$ tension. An interesting fact about {the set of equations describing the cosmic evolution of} dissipative fluids is their equivalence with {those when} matter-creation effects {arise} within homogeneous spacetimes, as discussed in Ref. \cite{zimdahl} from the kinetic theory perspective, see also Refs. given in \cite{equiv}. In this case the particle production is described by the particle creation pressure, $p_{c}$, which in turn is characterized by a $\Gamma$ term. Such pressure plays the role of $\Pi$ in equations (\ref{eq:pressure}), (\ref{eq:energy}) and (\ref{eq:accel}), i.e., we now write $p_{\mathrm{eff}}=p+p_{c}$ for the effective pressure and from this expression we can observe, $p_{c} \leftrightarrow \Pi$. From now on we will characterize the deviations from equilibrium pressure simply as $\Pi$ for simplicity in the notation. For the matter creation scenario we must also consider that the particle number is non conserved
\begin{equation}
    \dot{n}+3Hn = n\Gamma,
    \label{eq:number}
\end{equation}
then $\Gamma > 0$ represents production of particles. This is an important difference with dissipative processes where the particle number is conserved ($\Gamma=0$). Now we comment about the thermodynamics implications of both models. The Gibbs equation reads
\begin{equation}
    Tds = d\left(\frac{\rho}{n}\right) + pd\left(\frac{1}{n}\right),\label{eq:gibbs}
\end{equation}
where $s$ is the entropy per particle. The time derivative for the entropy is written from the above equation, one gets
\begin{equation}
    nT\dot{s} = -(\rho+p)\Gamma -3H\Pi.
\end{equation}
where we have considered Eqs. (\ref{eq:energy}) and (\ref{eq:number}). If the particle number is conserved, $\Gamma=0$, we recover the standard expression for entropy production within the dissipative scenario, $nT\dot{s} = - 3H\Pi$ \cite{maartens}. On the other hand, for matter creation effects the adiabatic condition, $\dot{s}=0$, is useful to relate the quantities $\Gamma$ and $\Pi$, yielding
\begin{equation}
    \Pi = - \frac{(\rho + p)}{3H}\Gamma, \label{eq:adiabatic}
\end{equation}
if created matter behaves as dark matter then we must consider $p=0$ in the previous equation. Some comments are in order, in the dissipative context $\Pi$ must be negative to guarantee positive production of entropy. From the above result we observe that the creation pressure is negative, therefore such effects are expected to contribute to the cosmic expansion; as can be seen, Eq. (\ref{eq:adiabatic}) is a special case due to the adiabatic condition. However, the covariant condition $s^{a}_{;a} = 0$ leads to $s^{a}_{;a} = n\dot{s} + sn\Gamma$, therefore we have entropy production from the $\Gamma$ term contribution even under the adiabatic condition, in other words, the increasing number of particles in the fluid induces the production of entropy, notice that the case $\Gamma<0$ is not physically consistent. The adiabatic condition only ensures that particles are accessible to a perfect fluid description immediately after created, this interpretation is discussed in detail in Refs. \cite{zimdahl, zimdahl2}.  In this sense, the deviations from local equilibrium pressure discussed previously lead to a more consistent description of the cosmic fluid from the thermodynamics point of view since entropy production is allowed. 

\subsection{Decaying dark matter plus dark energy}

Now we discuss some aspects of the decaying dark matter model, such decay is described by the following set of equations for a FLRW metric
\begin{align}
& \dot{\rho}_{\mathrm{m}} + 3H\rho_{\mathrm{m}} = - Q, \label{eq:ddm1}\\
& \dot{\rho}_{\mathrm{dr}} + 3H(1+\omega_{\mathrm{dr}})\rho_{\mathrm{dr}} = Q, \label{eq:ddm2}
\end{align}
where $\rho_{\mathrm{m}}$ denotes the dark matter energy density and $\rho_{\mathrm{dr}}$ denotes the dark radiation generated by the dark matter decay. In this case we have considered a barotropic EoS, $p_{\mathrm{i}}=\omega_{\mathrm{i}}\rho_{\mathrm{i}}$, where the subscript $i$ denotes different components, notice that we are restricting ourselves to the cold dark matter case, i.e. $\omega_{\mathrm{m}}=0$. According to Eqs. (\ref{eq:ddm1}) and (\ref{eq:ddm2}), this kind of model is related to the scenario of an interacting dark sector where the $Q$-term is a function that mediates the exchange of energy between dark species with the direction of the energy flow defined by the sign of $Q$. The interacting scheme represents a generalized cosmological model since the entropy production is allowed due to the presence of the $Q$-term; this differs from the adiabatic cosmic expansion obtained from the cosmological standard model; the interacting model also gives rise to a robust thermodynamic description of the cosmic fluids since some variable temperatures can be associated to the dark sector. For interesting reviews on this topic, see Refs. given in \cite{pavon}. In this case, $Q$ denotes the decay rate of dark matter. In general the form of the decay rate is given by, see for instance \cite{das1, das2}
\begin{equation}
    Q = \Delta \rho_{\mathrm{m}},
\end{equation}
%\footnote{\aal{Esta propuesta es equivalente a hacer una modificacion de la ecuación de estado de la materia oscura. Difiere de otros modelos de materia oscura decayente pues seria como tomar una tasa de decaimiento variable (proporcional a H(t)) mientras que en los modelos estandar es un valor fijo.}}
where $\Delta = \alpha H$, being $\alpha$ a positive constant. If $\alpha \sim 1$, then the mean lifetime of dark matter is about a Hubble time at different epochs. For the dark energy sector we consider the following continuity equation \cite{wcdm}
\begin{equation}
    \dot{\rho}_{\mathrm{de}} + 3H(1+\omega_{\mathrm{de}})\rho_{\mathrm{de}} = 0, \label{eq:de}
\end{equation}
where $\omega_{\mathrm{de}}$ is a constant parameter of state associated to dark energy {which is left as a free parameter in order to track effects on its values due to the occurrence of a irreversible process such as dark matter decay}. As can be seen from the previous equations, the total energy density is conserved, while those related to specific species do not. Adopting the standard relationship between the scale factor $a$ and the red-shift $z$ given as $1+z = a^{-1}$, we can perform a change of variable in the system of equations (\ref{eq:ddm1}) and (\ref{eq:ddm2}), then by considering the previous expressions for the decay rate we can obtain the following analytic solutions for the energy densities for both components \cite{das1, das2}
\begin{eqnarray}
\rho_{\mathrm{m}}(z) &=& \rho_{\mathrm{m,0}}(1+z)^{3+\alpha}, \label{eq:ddm3}\\
\rho_{\mathrm{dr}}(z) &=& \rho_{\mathrm{dr,0}}(1+z)^{3(1+\omega_{\mathrm{dr}})} + \rho_{\mathrm{m,0}}(1+z)^{3}\frac{\alpha}{\alpha-3\omega_{\mathrm{dr}}}\times \nonumber \\ &\times & \left[(1+z)^{3\omega_{\mathrm{dr}}}-(1+z)^{\alpha}\right]. \label{eq:ddm4}
\end{eqnarray}   
Notice that for $\alpha=0$ the standard behavior for the dark matter sector and radiation is recovered together with $\omega_{\mathrm{dr}}=1/3$. On the other hand, if $\omega_{\mathrm{dr}}=1/3$, the model is parametrized by a single parameter $\alpha$. For the dark energy sector one gets from Eq. (\ref{eq:de})
\begin{equation}
    \rho_{\mathrm{de}}(z) = \rho_{\mathrm{de,0}}(1+z)^{3(1+\omega_{\mathrm{de}})},
\end{equation}
being $\rho_{\mathrm{i,0}}$ the value of energy density at present time, $\rho_{\mathrm{i}}(z=0)=\rho_{\mathrm{i,0}}$. Then, for this cosmological model the Friedmann constraint reads $3H^{2}(z) = (\rho_{\mathrm{m}}(z)+\rho_{\mathrm{dr}}(z)+\rho_{\mathrm{de}}(z))$ and defining the normalized Hubble parameter, $E(z):=H(z)/H_{0}$ with $H_{0}$ being the Hubble constant given as $H(z=0):=H_{0}$, we can write
\begin{widetext}
\begin{equation}
    E(z) = \sqrt{\frac{\Omega_{\mathrm{m,0}}}{1-\alpha}(1+z)^{3+\alpha}+\left(\Omega_{\mathrm{r,0}}+\frac{\alpha}{\alpha-1}\Omega_{\mathrm{m,0}}\right)(1+z)^{4} + \Omega_{\mathrm{de,0}}(1+z)^{3(1+\omega_{\mathrm{de}})}}, \label{eq:1stmodel}
\end{equation}
\end{widetext}
where $\Omega_{\mathrm{i}}(z):=\rho_{\mathrm{i}}(z)/3H^{2}_{0}$ are the fractional energy densities and we have considered the standard parameter state for dark radiation, $\omega_{\mathrm{dr}}=1/3$. For $z=0$ in Eq. (\ref{eq:1stmodel}) we obtain $\Omega_{\mathrm{m,0}}+\Omega_{\mathrm{r,0}}+\Omega_{\mathrm{de,0}} =1$, which is the usual normalization condition. 

\subsection{Modified $\Lambda$CDM model}

Following the line of reasoning of Ref. \cite{velten}, we now proceed to explore the degeneracy problem at the background level between the cosmological model (\ref{eq:1stmodel}) and a $\Lambda$CDM model with a modified dark matter sector. This means that if both cosmological models give rise to the same expansion for the universe, then we must have $E(z)=E_{\Lambda}(z)$. Consequently, both parameter spaces can be mapped to each other. The modification made to the dark matter sector of the $\Lambda$CDM model is given by the consideration of an effective non-perfect pressure $p_{\mathrm{eff}}$ instead of the local equilibrium pressure, and then in our analysis the $\Pi$ term could be associated to an effective viscous pressure or a matter creation pressure. We first focus on the viscous scenario. Under the considerations commented above the pressure of matter sector takes the form
\begin{equation}
    p_{\mathrm{eff,vm}} = p_{\mathrm{m}}+\Pi = \Pi,
\end{equation}
since $p_{\mathrm{m}}=0$ for dark matter. Notice that in this case $\omega_{\mathrm{eff,vm}}\neq 0$. Thus we have 
\begin{equation}
    E_{\Lambda}(z) = \sqrt{\Omega_{\mathrm{vm}}(z)+\bar{\Omega}_{\mathrm{r,0}}(1+z)^{4}+\Omega_{\Lambda}},\label{eq:modif}
\end{equation}
where the viscous matter must be determined from the continuity Eq. (\ref{eq:energy})
\begin{equation}
    (1+z)\frac{d\Omega_{\mathrm{vm}}}{dz}+4\bar{\Omega}_{\mathrm{r}}=2(1+z)E_{\Lambda}\frac{dE_{\Lambda}}{dz}, \label{eq:viscous}
\end{equation}
we have used Eq. (\ref{eq:accel}) to write the viscous pressure as $\Pi =\rho_\Lambda-\frac{1}{3}\rho_r-3H^{2}-2\dot{H}^{2}$. From the degeneracy condition $E(z)=E_{\Lambda}(z)$, one gets the following expression for viscous dark matter. 
\begin{widetext}
\begin{equation}
    \Omega_{\mathrm{vm}}(z) = \frac{\Omega_{\mathrm{m,0}}}{1-\alpha}(1+z)^{3+\alpha}+\left(\Omega_{\mathrm{r,0}}+\frac{\alpha}{\alpha-1}\Omega_{\mathrm{m,0}}-\bar{\Omega}_{\mathrm{r,0}}\right)(1+z)^{4} + \Omega_{\mathrm{de,0}}(1+z)^{3(1+\omega_{\mathrm{de}})}-\Omega_{\Lambda}.
\end{equation}
\end{widetext}
Therefore, if we insert the result given above in Eq. (\ref{eq:viscous}), we get the following expression.
\begin{widetext}
\begin{equation}
     \left(\frac{3+\alpha}{1-\alpha}\right)\Omega_{\mathrm{m,0}}(1+z)^{3+\alpha}+4\left(\Omega_{\mathrm{r,0}}+\frac{\alpha}{\alpha-1}\Omega_{\mathrm{m,0}}-\bar{\Omega}_{\mathrm{r,0}}\right)(1+z)^{4}+3\Omega_{\mathrm{de,0}}(1+\omega_{\mathrm{de}})(1+z)^{3(1+\omega_{\mathrm{de}})}+4\bar{\Omega}_{\mathrm{r,0}}(1+z)^{4} = 2(1+z)E_{\Lambda}\frac{dE_{\Lambda}}{dz}. \label{eq:conds}
\end{equation}
\end{widetext}
If we consider the conditions given in \cite{velten} we have $\Omega_{\mathrm{r,0}}=\bar{\Omega}_{\mathrm{r,0}}$ since the density of the radiation fluid is obtained from the temperature of the CMB, in other words, this parameter is not model dependent. Besides, if the correct amount of matter is established correctly from observations then $\Omega_{\mathrm{m,0}}=\Omega_{\mathrm{vm,0}}$, which in turn implies that the dark energy density parameter will be also well determined for any model, then $\Omega_{\mathrm{de,0}}=\Omega_{\Lambda}$. Therefore, Eq. (\ref{eq:viscous}) takes the form
\begin{widetext}
\begin{equation}
    (1+z)\frac{d\Omega_{\mathrm{vm}}}{dz}=\frac{\Omega_{\mathrm{vm,0}}}{1-\alpha}(1+z)^{3}\left[(3+\alpha)(1+z)^{\alpha}-4\alpha(1+z) \right]+3\Omega_{\mathrm{de,0}}(1+\omega_{\mathrm{de}})(1+z)^{3(1+\omega_{\mathrm{de}})}, \label{eq:viscous2}
\end{equation} 
\end{widetext}
where the r.h.s. of Eq. (\ref{eq:viscous}) is replaced by the l.h.s. of Eq. (\ref{eq:conds}). Notice that the viscous matter sector can be written in terms of the parameters of the cosmological model (\ref{eq:1stmodel}). It is worth mentioning that the solution for $\Omega_{\mathrm{vm}}(z)$ does not depend on a specific election of $\Pi$, then this solution would be valid to characterize viscous effects obtained from the full theory of the causal Israel-Stewart formalism \cite{is} and even from the non-linear extension formalism given in \cite{nonlinear}, which is more adequate to describe an expanding fluid. Since matter creation effects are uniquely characterized by deviations from the equilibrium pressure, Eq. (\ref{eq:viscous2}) can also describe the created matter scheme, $\Omega_{\mathrm{vm}}(z) \longrightarrow \Omega_{\mathrm{cm}}(z)$, both scenarios will provide a solution for the matter sector in terms of the parameters associated with the decaying dark matter model plus dark energy $(\alpha,\omega_{\mathrm{de}})$ given in Eq. (\ref{eq:1stmodel}). In the context of the matter creation scenario, the Ansatz for the $\Gamma$ term is necessary to determine the form of the creation pressure. However, as can be seen in our description, we can dispense of the explicit form of this term. See for instance Ref. \cite{gamma} for examples of $\Gamma$.\\

By solving Eq. (\ref{eq:viscous2}) we can finally write a precise expression for the Hubble parameter of the modified $\Lambda$CDM model (\ref{eq:modif}), yielding 
\begin{widetext}
\begin{equation}
  E_{\Lambda}(z) = \sqrt{\frac{\Omega_{\mathrm{vm,0}}}{1-\alpha}(1+z)^{3+\alpha}+\Omega_{\mathrm{de,0}}(1+z)^{3(1+\omega_{\mathrm{de}})}+\left( \Omega_{\mathrm{r,0}}-\alpha \frac{\Omega_{\mathrm{vm,0}}}{1-\alpha}\right)(1+z)^{4}}.\label{eq:modif2}   
\end{equation}
\end{widetext}
This latter expression will be tested against observations to determine the cosmological implications of considering deviations from the equilibrium pressure within the matter sector under the degeneracy problem perspective. The usual normalization condition for the cosmological parameters is obtained from Eq. (\ref{eq:modif2}) at $z=0$ as expected.

We would like to highlight the fact that, under the inclusion of viscous effects and the consideration of the degeneracy condition at the background level between both models, the expansion of the universe within the viscous $\Lambda$CDM  model (\ref{eq:modif2}) behaves exactly as in the decaying dark matter model given in (\ref{eq:1stmodel}). However, the physical interpretation is distinct for each case; as can be seen, the modified $\Lambda$CDM model has a matter sector with a viscous pressure that can be modeled by means of the decaying dark matter model parameters (as we will see below) and the nature of the CC changes into a dynamical behavior, $\Lambda \rightarrow \Lambda(z)$, characterized by the parameter of state $\omega_{\mathrm{de}}$.\\ 
%This construction indicates that any viscous contribution considered within the $\Lambda$CDM model for the matter sector could be mapped to a decaying dark matter scenario.\\

\begin{center}
\begin{figure}[htbp!]
    \includegraphics[width=15.5cm,height=7cm]{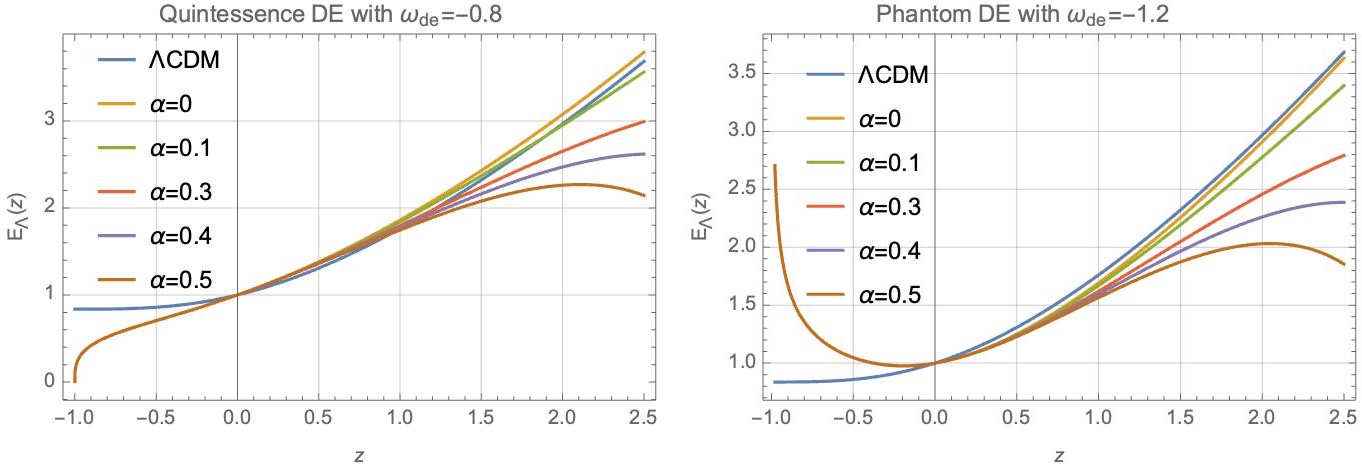}
    \caption{Behavior of the parameter $E_{\Lambda}(z)$. For comparison purposes, we use the values $\Omega_{\mathrm{r,0}}=0$, $\Omega_{\mathrm{m,0}}=0.3$, $\Omega_{\mathrm{de,0}}=0.7$. We focus on the recent evolution of the universe $z\approx 0.6$, since for that redshift value the transition decelerated-accelerated for the cosmic expansion is expected to take place \cite{moresco}.}
    \label{fig:hubbleparam}
\end{figure}
\end{center}

In Fig. (\ref{fig:hubbleparam}) we show the normalized Hubble parameter (\ref{eq:modif2}) and compare it to the $\Lambda$CDM model for the recent evolution of the universe. As can be seen, in the past, the cases are practically indistinguishable for a small value of $\alpha$. However, as the value of $\alpha$ increases, we can observe deviations from $\Lambda$CDM in the past for $z>1$. Some other differences with respect to $\Lambda$CDM arise in the far future ($z=-1$), whereas the Hubble parameter for $\Lambda$CDM tends to the constant value given by, $\sqrt{\Omega_{\Lambda}}$, in our model it vanishes for quintessence and diverges for phantom dark energy, therefore in this phantom scenario the singularity is {\it kicked away}, this is known as {\it little rip} scenario \cite{lr}, i.e., the model does not exhibit a true (big rip) singularity at a finite cosmic time, the singularity is only allowed until infinite time has elapsed. See, for instance \cite{lr2}, where it is discussed that viscous effects can naturally lead to a little rip cosmic evolution avoiding all problems of a big rip singularity; see also \cite{lr3}.\\ 

It is worthy to mention that for quintessence dark energy all values of $\alpha$ considered tend to increase the value of the normalized Hubble parameter w.r.t. the $\Lambda$CDM model in the low red-shift region, $0<z<1$, this is an interesting characteristic since such behavior is known to alleviate some tensions of $\Lambda$ CDM; see, for instance Ref. \cite{vagnozzi} where this kind of condition was studied for different dark energy models.      
\onecolumngrid
\twocolumngrid
To end this section we reconstruct the form of $\Pi$ from $\Pi =\rho_\Lambda-\frac{1}{3}\rho_r-3H^{2}-2\dot{H}^{2}$ and our previous results, yielding
\begin{align}
    & \Pi = -3H^{2}_{0}E^{2}_{\Lambda}(z)\left(1-\Omega_{\mathrm{de,0}}-\frac{1}{3}\Omega_{\mathrm{r,0}}(1+z)^{4}\right) \nonumber \\
    & + 2H^{2}_{0}(1+z)E_{\Lambda}\frac{dE_{\Lambda}}{dz}. \label{eq:pi}
\end{align}
Notice that the first term in the previous expression resembles the Eckart model, $\Pi=-3H_{0}E_{\Lambda}(z)\xi(z)$ with viscosity coefficient, $\xi(z) = H_{0}E_{\Lambda}(z)$ \cite{eckart}. However, the above result indicates that we are dealing with a more general description for viscous effects, and our construction does not depend on the usual and intricate constitutive equations which describe such effects; these equations depend on the framework used to introduce the viscosity effects in the cosmic fluid. According to the form of Eq. (\ref{eq:pi}) our construction is more general than the one considered in \cite{velten}, where the Eckart model was proved. It is worth mentioning that in this case the viscous pressure is reconstructed and characterized by the physical parameters $\alpha$, $\Omega_{i}$ and $H_{0}$ coming from the decaying dark matter model instead of the usual arbitrary parameters $\xi_{0}$ and $s$. In general, the election of the values for the parameters $\xi_{0}$ and $s$ is only justified and motivated by the mathematical simplification that can be obtained on the dynamical equation obeyed by the Hubble parameter in the viscous description. \\

In Fig. (\ref{fig:Pi}) we show the behavior of the viscous pressure (\ref{eq:pi}) as a function of the red-shift for different values of the decaying rate. In any case, we observe that $\Pi$ suffers a transition to an attractor behavior around $z\sim 0.25$. After this point (for $z \leq 0.25$) and throughout the future cosmic history, $\Pi$ evolves exactly the same way, independently of the decay rate value $\alpha$. At the far future, the viscosity pressure in the phantom scenario tends to become more negative as the universe evolves, whilst for the quintessence case it remains negative but close to zero. At times before the attractor transition, the behavior depends on the value of $\alpha$. Namely, for very small $\alpha$ values, $\Pi$ takes large values in the far past and decreases along the cosmic evolution for quintessence and phantom dark energy until today. In contrast, for larger values of the decaying rate ($\alpha > 0.1$) in both cases, $\Pi$ increases from negative values in the distant past, towards positive ones. For the phantom case, this transition occurs before than for the quintessence case. For $0.1 < \alpha \leq 0.6$ the positive phase of $\Pi$ in the past lasts a short while and finishes around the attractor transition. In the extreme case with $\alpha>0.8$ this phase does not occur. In addition, we computed the deceleration parameter (see figure (\ref{fig:q})) and observe that it exhibits a transition from positive to negative before the change of sign in $\Pi$. On the other hand, as the value of the parameter $\alpha$ increases, such a transition for $q$ is lost, and the deceleration parameter is always negative. This is interesting as the viscous effects give rise to an accelerated expansion even for the quintessence case, in which the state parameter of dark energy is not as negative as for a CC. Notice that since in our formulation we have $\Lambda \rightarrow \Lambda(z)$, the behavior of $q$ at late times deviates from the expected value $-1$ obtained for the CC due to the existence of the viscous contribution in the matter sector. However, as can be seen in the plot, such an attractor can be recovered by considering the appropriate value for $\omega_{\mathrm{de}}$.\\ 

Finally, using Eq. (\ref{eq:adiabatic}) with $p=0$ we can construct the $\Gamma$ term if we interpret $\Pi$ as the creation pressure, in this case we have
\begin{equation}
    \Gamma = -\frac{\Pi}{H_{0}E_{\Lambda}(z)\Omega_{\mathrm{vm}}(z)},
\end{equation}
where $\Omega_{\mathrm{vm}}(z)$ is obtained from Eq. (\ref{eq:viscous2}) and $\Pi$ can be taken from (\ref{eq:pi}). 
The source term obtained in this description deviates from the usual functional form used in the literature for $\Gamma$ \cite{gamma}, that is, $\Gamma \propto E^{m}(z)$, being $m$ a constant. In this case we observe, $\Gamma = \Gamma(E,E')$, where the prime denotes derivatives w.r.t. red-shift. This behavior is interesting since incorporates matter couplings with curvature, see, for instance, Ref. \cite{odintsov}. Given the form of $\Pi$, we observe that the particle production behaves in a different way for quintessence and phantom dark energy and will be well defined always that $\Gamma >0$ since the particle production is possible due to the transfer of energy from the gravitational field to matter \cite{creation}.
%This means that in both scenarios, the contribution of viscous effects  in order to have a positive cosmic acceleration, becomes relevant.
\onecolumngrid
\begin{center}
\begin{figure}[htbp!]\includegraphics[width=18.5cm,height=7.5cm]{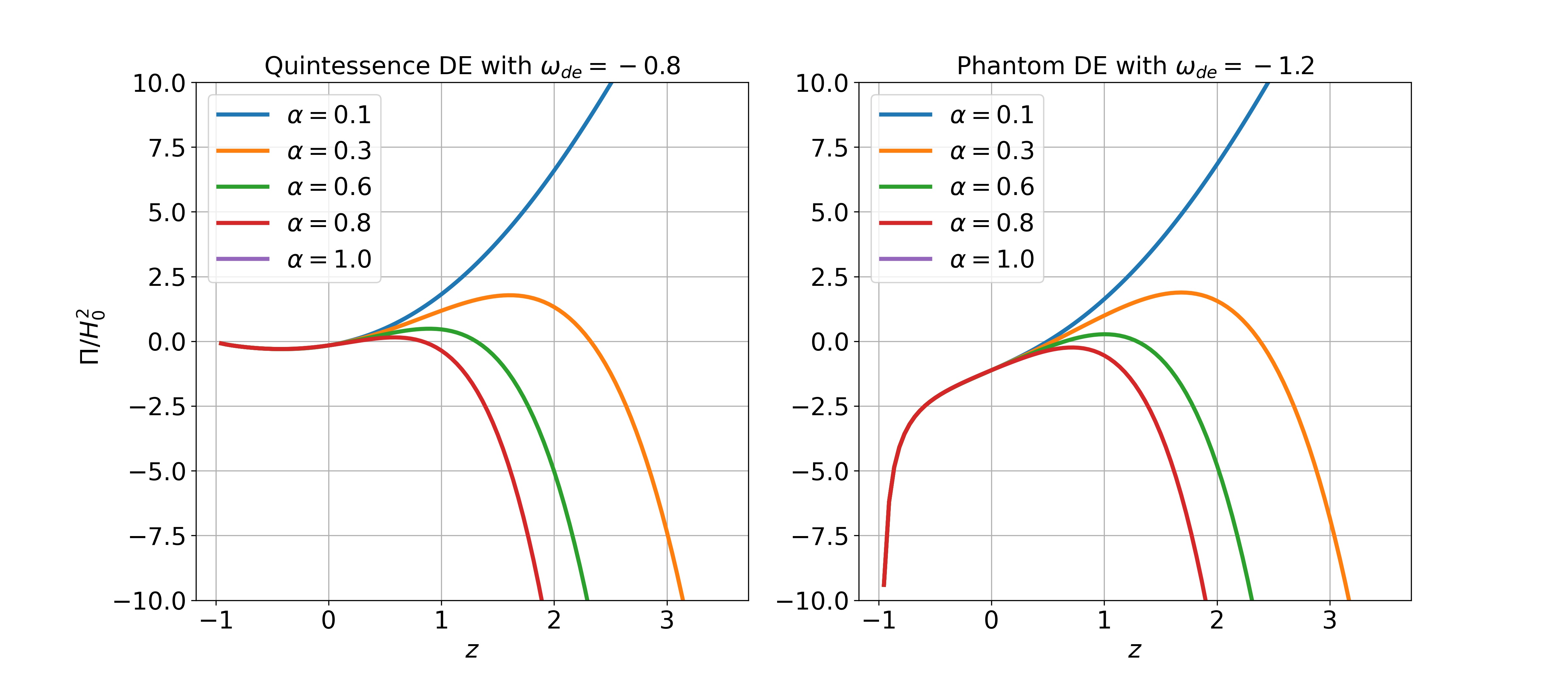}
        \caption{Reconstructed viscous pressure $\bar{\Pi}:=\Pi/H_{0}^{2}$ given in Eq. (\ref{eq:pi}).}
    \label{fig:Pi}
\end{figure}

\begin{figure}[htbp!]
    \includegraphics[width=18.5cm,height=7.5cm]{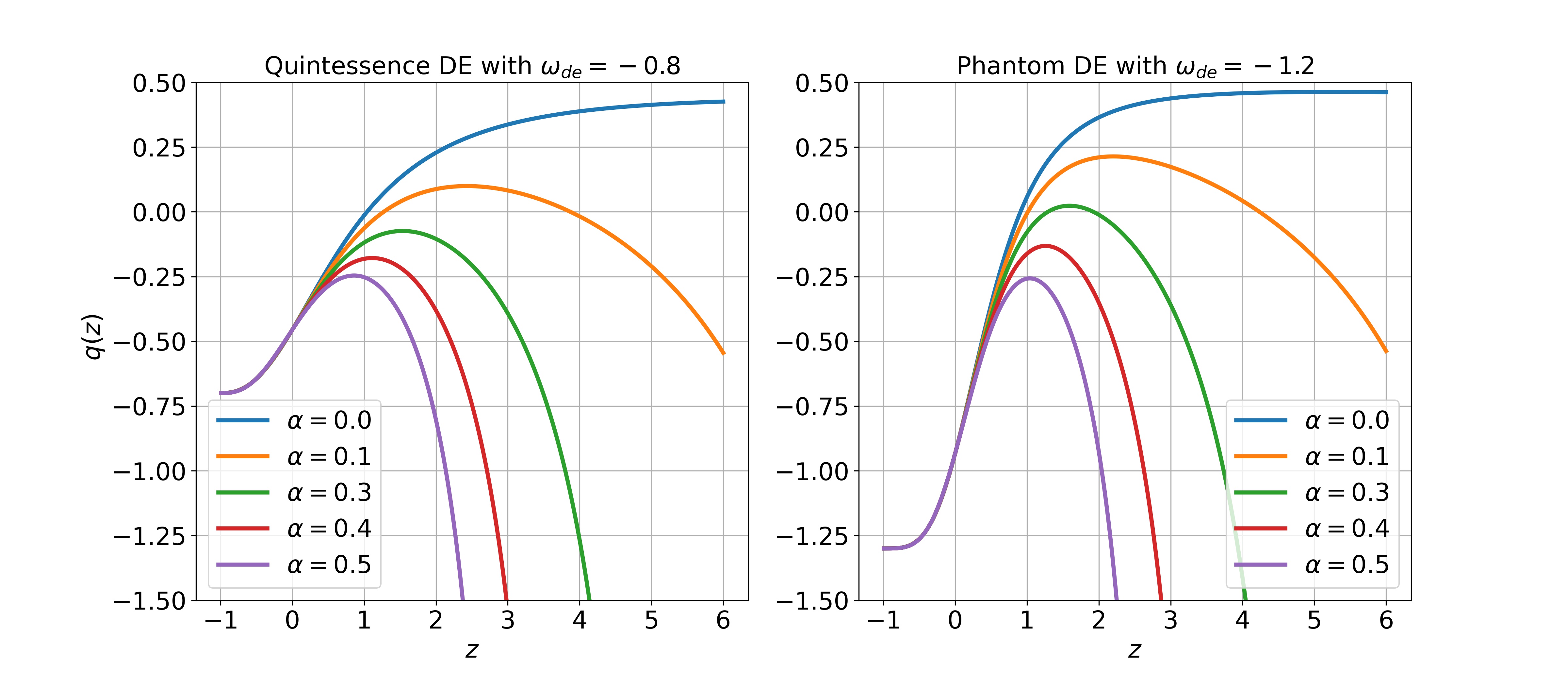}
        \caption{Behavior of the deceleration parameter $q$ defined as $q(z):=-1+(1+z)(E_{\Lambda})^{-1}dE_{\Lambda}/dz$.}
    \label{fig:q}
\end{figure}
\end{center}
\twocolumngrid  

%%%%%%%%%%%%%%%%%%%%%%%%%%%%%%%%%
\section{Viability of the model based on observations}
\label{sec:data}
%%%%%%%%%%%%%%%%%%%%%%%%%%%%%%%%%%%

\subsection{Test of the expansion of the universe with Supernovae IA from Union 2.2}
In this section we aim to test the capability of our model to predict the accelerated expansion of the universe. Firstly, in order to make a minimal test, we compare our theoretical predictions of luminous distance modulus as a function of redshift with measurements from the sample of supernovae type IA from the Supernovae Cosmology Project compilation Union 2.2 \cite{Suzuki:2012}. It is important to mention that in this work we limit ourselves to considering a small baseline sample of supernovae given that our purpose is to make a minimal test of the viability of our model at the level of the background cosmology and to obtain rough estimates for the parameters rather than determine precise Bayesian inferences of their values. Once this minimal test is made, in further work we aim to extend our analysis and strengthen the results presented here by considering, on one hand larger and more recent datasets such as Pantheon, DES, LIGO etc. \cite{PantheonScolnic:2021,DES:2024,LIGOScientific:2017}. On the other hand, by taking into account the test for the perturbations, it is worth mentioning that the marginalized distributions obtained here would be useful as priors to such a further analysis. In regard of the background cosmology, considering distance measurements of larger samples of supernovae at larger redshifts such as those of DES and Pantheon, constraints on $\alpha$ are expected to be stronger since the sensitivity of $\mu$ increases at high redshift. Including such data sets would be helpful to tighten constraints on the Hubble constant. Furthermore, comparing theoretical predictions for perturbations within this model with observations of LSS from different surveys would greatly enrich the analysis. For instance, effects of the viscosity on expansion of the universe could be distinguished from those on the structure formation process, specially those on small scales \cite{hu2,Lopez:2022}, and this would be useful to investigate the scale dependence of the $\alpha$ parameter. In order to obtain the theoretical prediction, we firstly computed the luminous distance as function of red-shift given as:
\begin{equation}
d_L(z) = a_0(1+z)S_k(\chi(z)),
\end{equation}
 as we are assuming an universe with a flat spatial geometry therefore: 
\begin{equation}
S_k(D_M(z))=D_M(z)=\int_0^z\,\frac{c\,dz}{H(z)}= \frac{c}{H_0}\int_0^z\,\frac{dz}{E_\Lambda(z)}.
\end{equation} 
Where $c$ is the speed of light and the denominator in the integrand corresponds to that given by (\ref{eq:modif2}). We carried out the corresponding numerical integrations by implementing the trapezoid method in our own Python code. Since the truncation error of this method is of order 2 of the step size, the accuracy of our results is good enough for our purposes. Afterwards we computed the distance modulus simply by using:
\begin{equation}
\mu(z) := m-M=5\log_{10}\left(\frac{d_L(z)}{1\,\mathrm{pc}}\right)-5.
\end{equation}

\begin{figure}[htbp!]
\includegraphics[width=9.5cm,height=7.5cm]{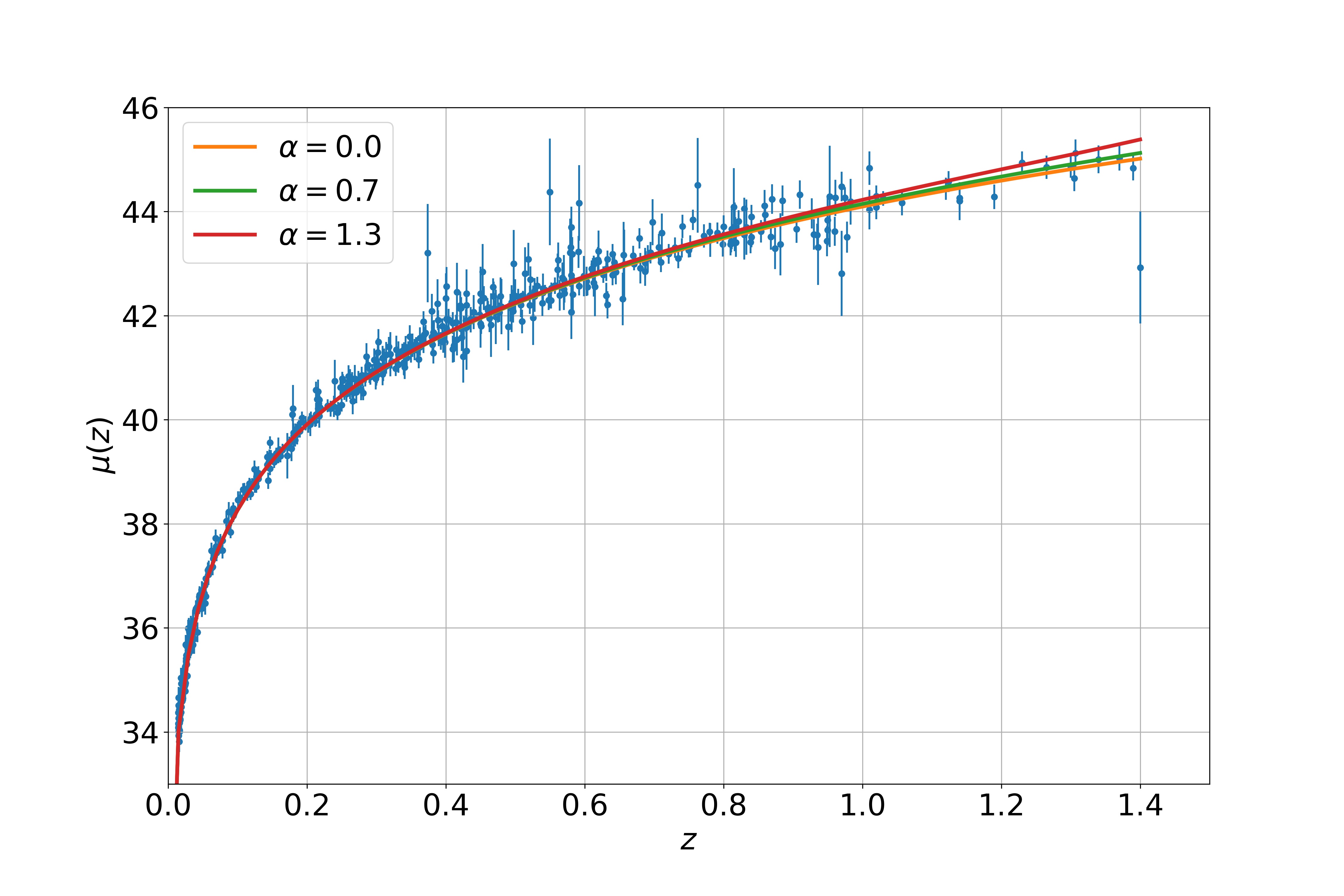}
\caption{Predicted distance modulus as function of red-shift from our model (purple line) overlapped to the observed modulus for Union 2.2 supernovae. The values of the parameters for this model corresponds to $\Omega_mh^2=0.126$, $H_0 =73\,\mathrm{km/s/Mpc}$, $\omega_0=-0.7$ and $\alpha=0.5$.}
\label{fig:Union}
\end{figure}
As shown in Figure (\ref{fig:Union}), as the decay rate parameter $\alpha$ increases, a larger distance modulus is expected and this modification is more pronounced at large red-shift values, while predictions for small red-shifts remain indistinguishable from those of $\Lambda$CDM. Given the small sensitivity of $\mu$ to variations of $\alpha$, it is expected that these data cannot strongly constrain this parameter.  

\subsection{Test of the expansion of the universe with BAO}

\begin{figure}[htbp!]
    \includegraphics[width=9.5cm,height=7.2cm]{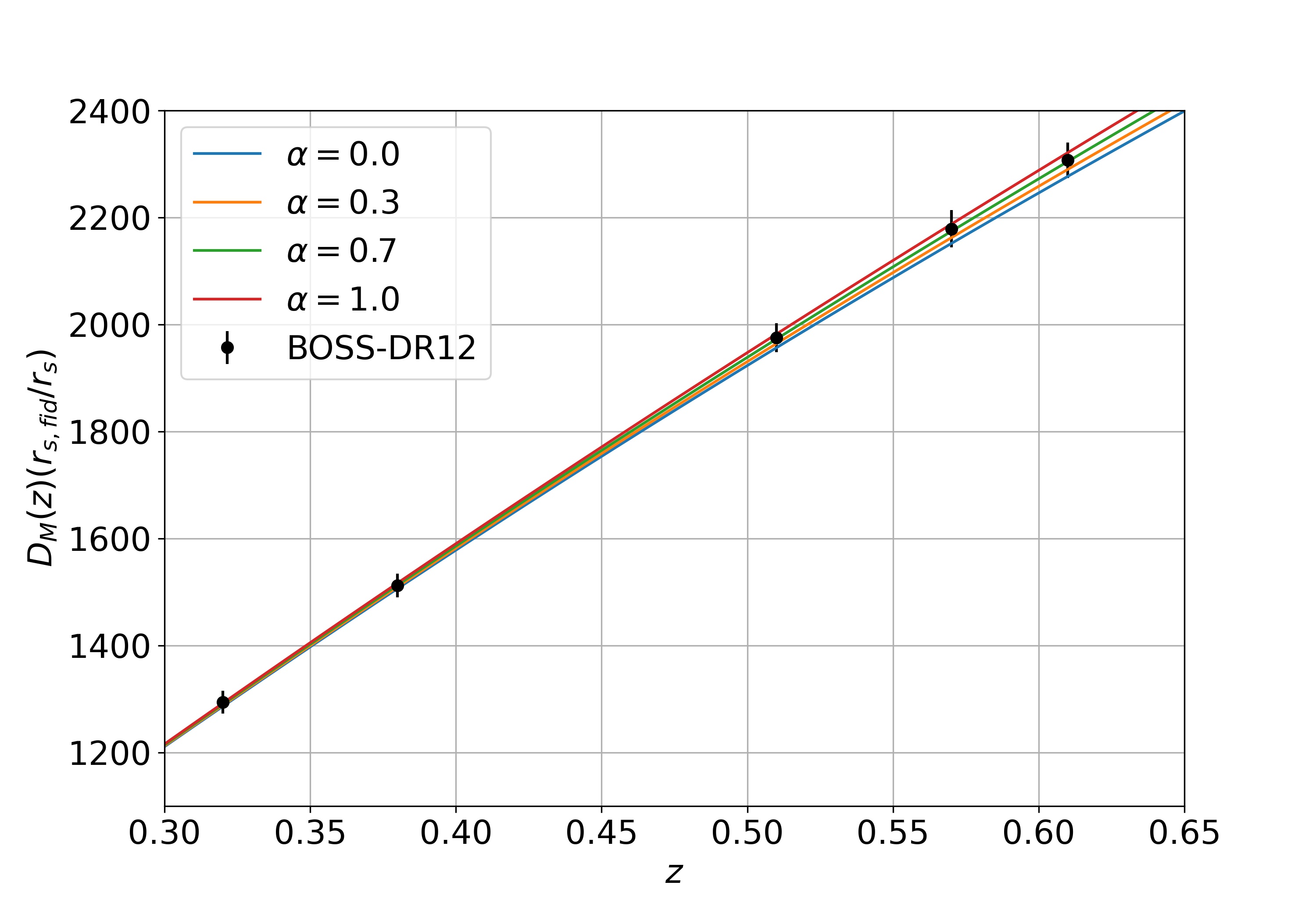}
     \includegraphics[width=9.5cm,height=7.2cm]{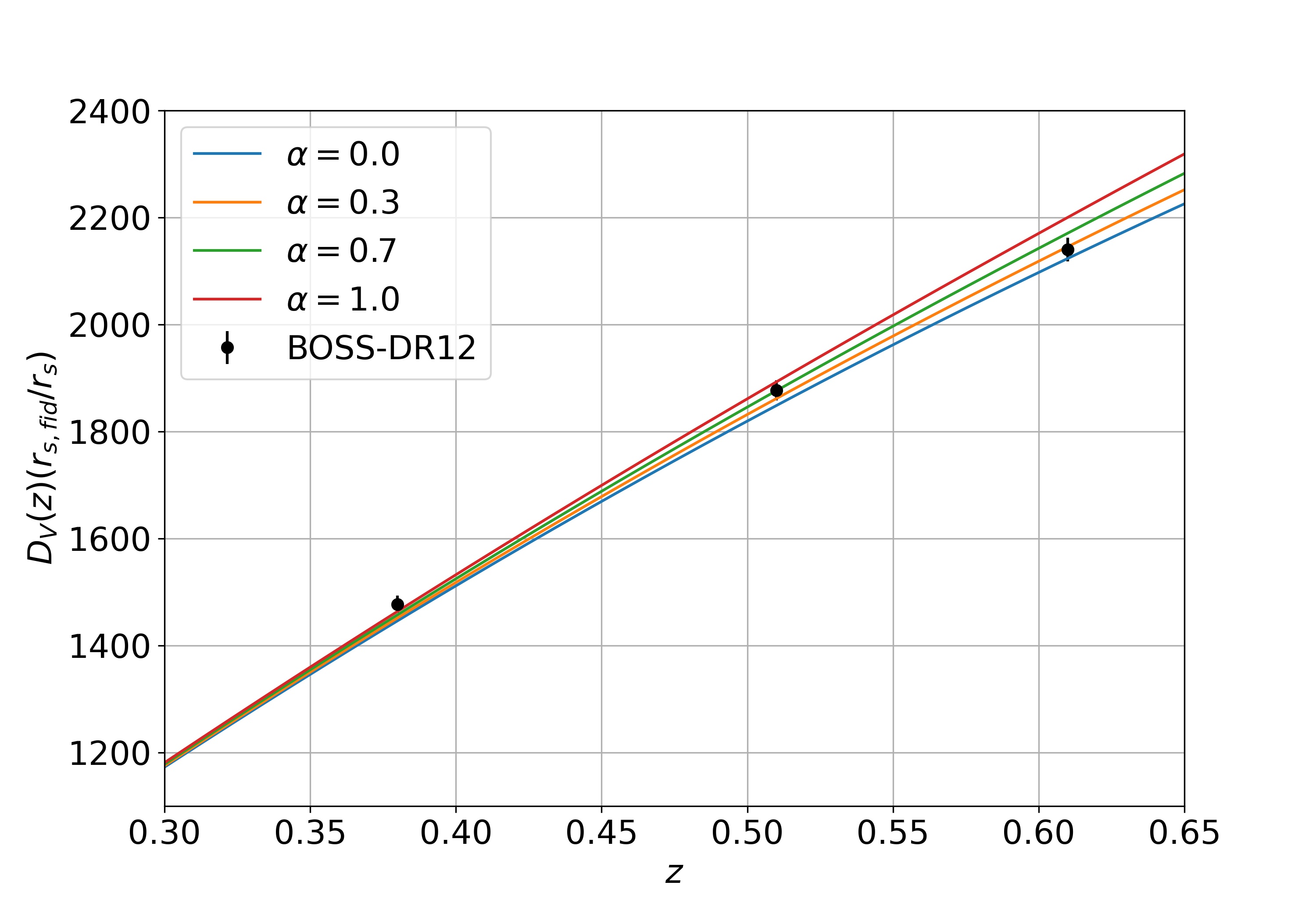}
\caption{(Solid lines) Comovil distance and the volume averaged angular diameter distance predicted in a sample of models with various values of $\alpha$. (Dots with errors) BAO distance measurements inferred from the 12th release of SDSS reported in \cite{Alam:2017}.}
\label{fig:DM}
\end{figure}

Given that SNIa distance measurements are not sensitive to variations of $\alpha$, constraints for the parameters are expected to be weak and that degeneration between them exists, therefore, in order to overcome such inconveniences, we additionally consider in our analysis that measurements of the BAO feature along the line of sight and transverse directions can separately measure $H(z)$ and the comoving angular diameter distance $D_M(z)$ from SDSS-III data \cite{Alam:2017}. Variations in the cosmological parameters or the pre-recombination energy density can alter the sound horizon of acoustic oscillations given by:
\begin{equation}
r_s(z)=\frac{c}{H_0}\int_{z^*}^\infty \frac{dz}{\sqrt{3(1+\mathcal{R})}E_\Lambda(z)},    
\end{equation}
where $\mathcal{R}=\frac{3\rho_b(z)}{4\rho_\gamma(z)}=\frac{3\Omega_{b0}}{4\Omega_{\gamma0}}(z+1)^{-1} $ and $z^*$ corresponds to the red-shift at recombination. Therefore, BAO measurements actually constrain the combinations of $D_M(z)/r_s$ and $H(z)r_s$, therefore we compare the predictions of $D_M(z)r_{s,fid}/r_s$ and $H(z)r_s/r_{s,fid}$ with their inferred values from the observations of BAO by BOSS-DR12, where $r_{s,fid}$ corresponds to the fiducial value of $\Lambda$CDM for the sound horizon given as $r_{s,fid} = 147.78\,\mathrm{Mpc}$ which is used as normalization constant. 
In Figure (\ref{fig:DM}) the predictions of $D_M(z)r_{s,fid}/r_s$ are shown for different models corresponding to a range of values of $\alpha$. Clearly, this quantity barely depends on $\alpha$ at low red-shift, on the contrary, it is sensitive to this parameter at high red-shift. Within this range $\alpha \sim 0.5$ seems to fit properly the observational data. On the other hand, an angle-averaged galaxy BAO measurement constrains the following combination \cite{Alam:2017}:
\begin{equation}
D_V(z)= \left[\frac{cz}{H_0}\frac{D_M(z)^2}{E_\Lambda(z)}\right]^{1/3}.
\end{equation}

\subsection{Bayesian Analysis}

\begin{center}
  \begin{figure}
  \includegraphics[width=0.7\textwidth]{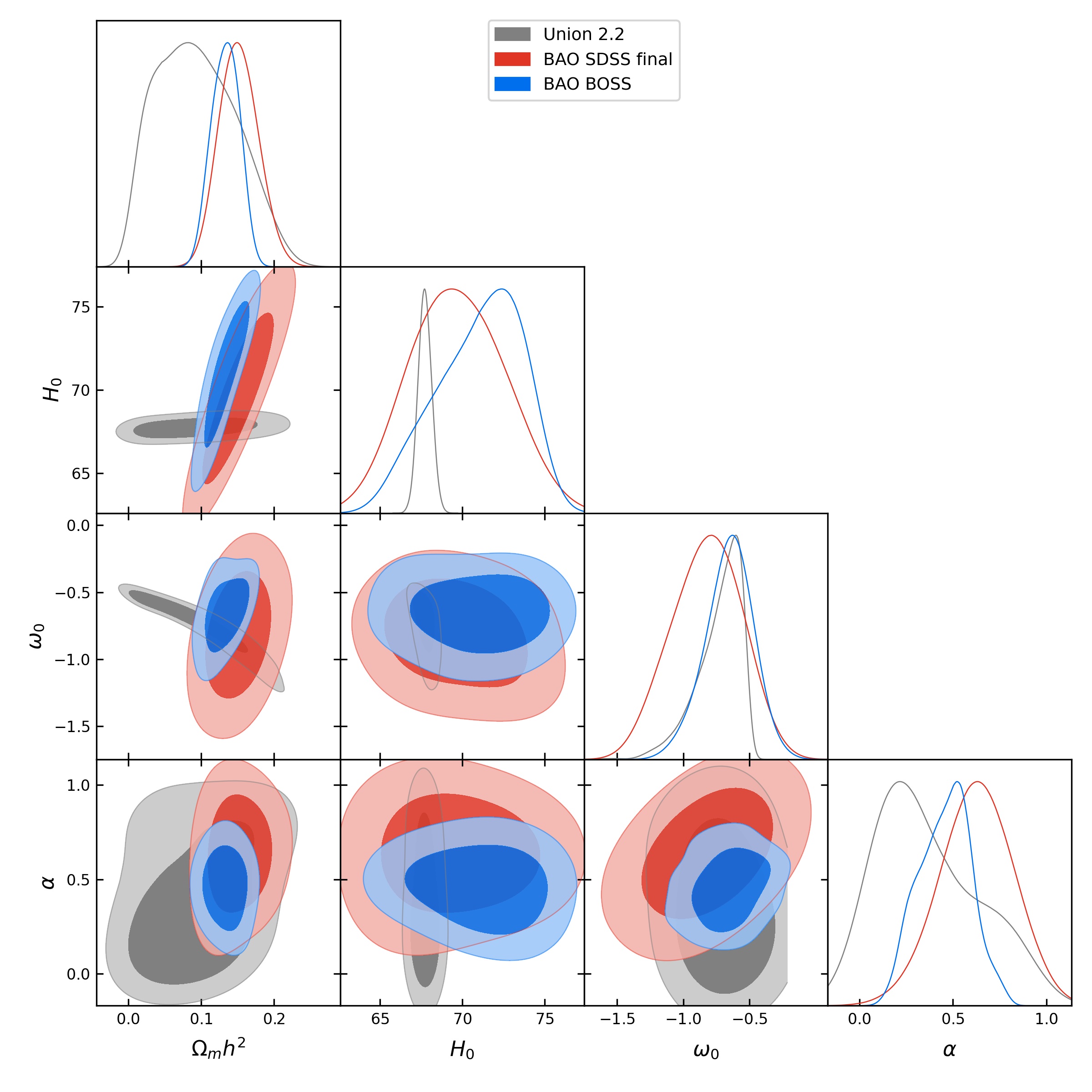}
\caption{1D and 2D marginalized posterior distributions for the parameters the viscous $\Lambda$CDM model.}
\label{fig:PD}
\end{figure}
\end{center}

In order to obtain baseline estimates for the parameters of the viscous $\Lambda$CDM model, we sample the corresponding parameter space using the Markov Chain Monte Carlo (MCMC) method. Specifically, we implemented the Metropolis-Hasting algorithm in our own Python code to construct a set of Markov chains that serve as samples in the space of parameters of our model used to determine the posterior distribution function corresponding to supernovae data from Union 2.2 and BAO data from BOSS-DR12 \cite{Suzuki:2012, Alam:2017}. Since we aim to obtain base estimations of the parameters of this model, we use flat priors within ranges that give rise to a physically admissible background evolution. Specifically, flat priors for parameters within the following ranges are considered: $\Omega_m h^2\,\in,\ (0,1)$, $\omega_0,\ \in ,\ (-2,0)$, $H_0,\ \in ,\ (30,120)$, $\alpha,\ \in ,\ (-2,2)$. Notice that we consider negative values for $\alpha$ which correspond to the case in which dark matter is produced by the inverse decay process; although positive values give rise to more realistic model, we take these cases into account for completeness and because the posteriors are independent of prior choice as desired. We marginalized the posterior distribution to estimate the best-fit parameters of our model and their uncertainty at the level of confidence $68\%$ and $95\%$ (see Figure (\ref{fig:PD}) and the Table (\ref{tab:bestfits}) at the end of this section).

% mcmc_file_name = "./modelo_alpha_union_Poissoniana.dat"
% mcmc_file_name2 = "./modelo_alpha_BAO_Gaussian.dat"
%% sample1=getdist.mcsamples.loadMCSamples('./Union_BAO', ini=None, jobItem=None, no_cache=False, settings={'ignore_rows':0.943}) 
%% sample2=getdist.mcsamples.loadMCSamples('./BAO_SDSS', ini=None, jobItem=None, no_cache=False, settings={'ignore_rows':0.74}) 
\onecolumngrid
%\begin{minipage}{0.5\textwidth}
  %  \includegraphics[width=0.9\textwidth]{Posteriors_Bao_DVr_Final.jpeg}
%\end{minipage}
\twocolumngrid
Interestingly, the marginalized distribution for $\alpha$ is pretty similar according to SNIa and BAO, with a best-fit value $\alpha_{bf} \simeq 0.5$. The fact that $\alpha \neq 0$ gives rise to a better fit of both datasets suggests that the occurrence of dissipative effects along the expansion history is not only feasible, but rather the dynamics of expansion might be better described in comparison to the standard $\Lambda$CDM. Another point that is important to stress is that marginalized distributions for $\Omega_mh^2$ and $H_0$ are consistent even at $1-\sigma$ for both datasets, SNIa and BAO. A slight tension can be observed for the dark energy state parameter $\omega_0$. Although this discrepancy in the estimate of $\omega_0$ for both datasets is not statistically significant, it can be due to the different features of the expansion of the universe at late and early times.  Also, it is worth noticing that estimates for these parameters according to BAO are stronger than those for SNIa, this can be explained by the large sensitivity of the theoretical function $D_M(z)$ to these parameters at the largest red-shift values in the sample and to the smallness of the error bars of the BAO measurements (see Figure (\ref{fig:DM})). It is also worth mentioning that the constrained values for the parameter $\omega_{0}$ are in good agreement with the bounds obtained by the DESI Collaboration for a dynamical dark energy model labeled as $\omega_{0} \omega_{a}$CDM \cite{desi}. 

To end this section we provide some comments on the consideration of the joint data in our analysis. For the flat priors mentioned above, we obtain the following best-fit values for the model parameters at the $68\%$ confidence level, {\boldmath$\Omega_m h^2$}: $0.1758^{+0.0042}_{-0.0012}$, {\boldmath$H_0$}: $77.83^{+0.34}_{-0.30}$, {\boldmath$\omega_0$}: $-1.197^{+0.077}_{-0.077}$ and {\boldmath$\alpha$}: $0.30^{+0.16}_{-0.21}$. As can be seen, the value of $H_0$ tends to high values in order to satisfy the strong constraint imposed by BAO on $\Omega_{m}h^2$. However, this tension is alleviated if the Gaussian prior, $\Omega_m h^2 = 0.1434^{+0.0020}_{+0.0020}$, obtained from the constraint of various CMB spectra at $68\,\%$ of confidence level is used in our analysis; see references given in \cite{planck1518} for the Planck Collaboration results. The best-fit values in the latter case are also shown in Table (\ref{tab:bestfits}). We highlight the fact that consideration of joint data together with the Gaussian prior for $\Omega_m h^2$ provides a better estimation of the parameter $\alpha$ w.r.t. SNIa and BAO datasets since the corresponding uncertainties decrease, and we also observe that the Hubble constant value is closer to that reported by the SH0ES Collaboration \cite{shoes}. The consideration of more datasets could improve the parameter estimations presented here.

\onecolumngrid
%\newpage
%\begin{widetext}
\begin{center}
\begin{table}
\begin{tabular}{| c | c | c | c |}
\hline
SNaI from Union 2.2 & BAO SDSS Final & BAO BOSS Fs  & Joint data\\
\hline
\begin{tabular} {   l  |  c  | c  }
\hline
Parameter &  68\% limits & 95\% limits\\
\hline
{\boldmath$\Omega_mh^2 $} &  $0.093^{+0.045}_{-0.069} $   & $0.093^{+0.10}_{-0.093} $ \\

{\boldmath$H_0 $} & $67.74^{+0.41}_{-0.41}$ & $67.74^{+0.82}_{-0.80}$\\

{\boldmath$\omega_{0} $} &$-0.730^{+0.21}_{-0.083}$ & $-0.73^{+0.26}_{-0.36}$ \\

{\boldmath$\alpha$} &$0.36^{+0.21}_{-0.36}$ & $0.36^{+0.58}_{-0.47}$ \\
\hline
\end{tabular}
 &  
\begin{tabular} {    c  | c }
\hline
  68\% limits& 95\% limits\\
\hline
$0.151^{+0.025}_{-0.025} $& $0.151^{+0.053}_{-0.051}$\\

$69.7^{+2.6}_{-2.6}$ & $69.7^{+5.6}_{-5.5}$\\

$-0.82^{+0.29}_{-0.26} $ & $-0.82^{+0.49}_{-0.52} $\\

 $0.63^{+0.19}_{-0.17}$ & $0.63^{+0.39}_{-0.39}$\\
\hline
\end{tabular}
&  
\begin{tabular} {    c  | c }
\hline
  68\% limits& 95\% limits\\
\hline
$0.134^{+0.019}_{-0.019}$ & $0.134^{+0.033}_{-0.036}$\\

$71.1_{-2.1}^{+3.2}$ & $71.1^{+4.4}_{-5.3}$\\

$-0.65^{+0.20}_{-0.17} $ & $-0.65^{+0.34}_{-0.40} $\\

$0.46^{+0.15}_{-0.14}$ & $0.46^{+0.26}_{-0.27}$\\
\hline
\end{tabular}
&
\begin{tabular} {  | c}
\hline
  68\% limits\\
\hline
$0.1240^{+0.0010}_{-0.00049}$\\

$75.85^{+0.27}_{-0.24}$\\

$-0.889^{+0.035}_{-0.040}$\\

$0.150^{+0.058}_{-0.090}$\\
\hline
\end{tabular}
\end{tabular}
\caption{Values for the best-fit parameters of the model at the level of confidence $68\%$ and $95\%$ and $68\%$ for the joint data analysis.}
    \label{tab:bestfits}
\end{table}
\end{center}
%\end{widetext}
\twocolumngrid

%%%%%%%%%%%%%%%%%%%%%%%%%%%%%%%%%%
\section{Final Remarks}
\label{sec:final}
%%%%%%%%%%%%%%%%%%%%%%%%%%%%%%%%%%%

In this work, we study a cosmological model in which a generic candidate of dark matter undergoes an exponential decay into relativistic dark particles at the late stages of evolution of the universe. Since the decay process of dark matter occurs out of chemical equilibrium given that the average kinetic energy is lower than the mass of the candidate, the entropy grows considerably and therefore it is irreversible. As a consequence, the fluid turns out to be not perfect, and a bulk viscosity pressure can be associated to it.\\ 

On the other hand, it is well known that different cosmologies describe identically the background evolution of the universe, that is, the solutions for the scale factor are degenerate between models. Therefore, by virtue of this degeneracy at the background level, a mapping between the space of parameters of different models can be established. In this work, we map the parameters of the cosmological model including decaying dark matter described above and a slight extension of the standard cosmological model in which dark matter corresponds to an effective viscous fluid and dark energy corresponding to a perfect fluid with barotropic equation of state. By setting this mapping, we are able to compute an estimate of the bulk viscosity of decaying dark matter in an effective manner and to describe the expansion of the universe according to this framework. As mentioned in the work, we restricted ourselves to background analysis. Given that the matter sector evolves in time in a different way in both models, some differences such as distinct moments of the matter-radiation equality ($z_{\mathrm{eq}}$) and the time of the last scattering ($z_{\mathrm{LSS}}$) are expected between the models. However, due to the degeneracy condition, the matter density could be adjusted in order to recover the same values for $z_{\mathrm{eq}}$ and $z_{\mathrm{LSS}}$ in the degenerate models. Therefore, the computation of spectrum of anisotropies in the cosmic microwave background and the matter power spectrum could exhibit a breaking of the degeneracy explored in this work at the background level. We leave this subject open for future investigation. As discussed previously, the reconstructed viscous pressure evolves in a different way depending on the value of the parameter $\omega_{\mathrm{de}}$. However, for phantom dark energy and quintessence, it has a negative contribution to the pressure of the dark matter sector from the recent past ($z\leq 1$) until the far future ($z=-1$). If we consider the same region for the deceleration parameter, we can observe a transition from decelerated to accelerated cosmic expansion for $\alpha \leq 0.6$. The behavior at the far future for this parameter deviates from the one expected in the CC scenario since in our case due to the mapping between models, we must reinterpret the CC as a variable dark energy. It is worthy to mention that despite the consideration of the phantom dark energy, $\omega_{\mathrm{de}} < -1$; the model does not exhibit a true (big rip) singularity; in this case the singularity is allowed until infinite time has elapsed; this is known as {\it little rip} and it has been explored previously within viscous cosmologies.\\   

In addition, by using Bayesian statistics, we test the expansion of the universe predicted by our model against observations of luminous distance and red-shift of type IA supernovae from the Union 2.2 sample and distance measurements from BAO detected in SDSS catalogs. By sampling the space of parameters of our model using the Markov-Chain-Monte-Carlo method implemented by means of the Metropolis-Hastings algorithm in our own Python code, we derive the corresponding 1D and 2D posterior distributions from which we obtain estimates of the model parameters and their uncertainties. As the main result of this analysis we found that dissipative effects along the cosmic history are feasible and could provide a better description of the cosmic expansion than the $\Lambda$CDM model.\\

In summary, a simple mapping at the background level between models in which processes out of equilibrium in the cosmic expansion at late times are allowed and minimally extended versions of the $\Lambda$CDM model can provide a natural and simple way to obtain reinterpretations of the cosmological concordance model in order to provide a more adequate description of the observable universe without invoke the existence of exotic components.  

\section*{Acknowledgments}
A.~A. and M.~C. work was partially supported by S.N.I.I. (CONAHCyT-M\'exico). G.~A.~P. was also supported by CONAHCyT through the program {\it Estancias Posdoctorales por México 2023(1)}. Anonymous reviewer is acknowledged for helpful suggestions.


\begin{thebibliography}{4}

\bibitem{planck1518}
P.~A.~R.~Ade, et al. (Planck Collaboration), Astron.\ Astrophys.\ {\bf 594}, A13 (2016); P.~A.~R.~Ade, et al. (Planck Collaboration), Astron.\ Astrophys.\ {\bf 571}, A16 (2014); N.~Aghanim, et~al. (Planck Collaboration), Astron.\ Astrophys.\ {\bf 641}, A1 (2020).

\bibitem{riess2024}
A.~G.~Riess, G.~S.~Anand, W.~Yuan, L.~M.~Macri, S.~Casertano, A.~Dolphin, L.~Breuval, D.~Scolnic, M.~Perrin and R.~I.~Anderson, arXiv:2401.04773 [astro-ph.CO].

\bibitem{hu}
W.~Hu, Astrophys.\ J. {\bf 506}, 485 (1998).

\bibitem{hu2}
M.~Kopp, C.~Skordis and D.~B.~Thomas, Phys.\ Rev.\ D {\bf 94}, 043512 (2016).

\bibitem{hu3}
M.~Kopp, C.~Skordis, D.~B.~Thomas and S.~Ili\'c, Phys.\ Rev.\ Lett. {\bf 120}, 221102 (2018).

\bibitem{large}
D.~Baumann, A.~Nicolis, L.~Senatore and M.~Zaldarriaga, J.\ Cosmol.\ Astropart.\ Phys. {\bf 07},
051 (2012).

\bibitem{takahashi}
K.~Ichiki, M.~Oguri and K.~Takahashi, Phys.\ Rev.\ Lett. {\bf 93}, 071302 (2004).

\bibitem{lesgourgues}
V.~Poulin, P.~D.~Serpico and J.~Lesgourgues, J.\ Cosmol.\ Astropart.\ Phys. {\bf 08}, 036 (2016).

\bibitem{das1}
O.~E.~Bjaelde, S.~Das and A.~Moss, J.\ Cosmol.\ Astropart.\ Phys. {\bf 10}, 017 (2012).

\bibitem{das2}
K.~L.~Pandey, T.~Karwal and S.~Das, J.\ Cosmol.\ Astropart.\ Phys. {\bf 07}, 026 (2020).

\bibitem{bringmann}
T.~Bringmann, F.~Kahlhoefer, K.~Schmidt-Hoberg and P.~Walia, Phys.\ Rev.\ D {\bf 98}, 023543 (2018).

\bibitem{vattis}
K.~Vattis, S.~M.~Koushiappas and A.~Loeb, Phys.\ Rev.\ D {\bf 99}, 121302 (2019).

\bibitem{bh}
T.~Nakamura, M.~Sasaki, T.~Tanaka and K.~S.~Thorne, Astrophys.\ J. {\bf 487}, L139 (1997); M.~Raidal, V.~Vaskonen and H.~Veermae, J.\ Cosmol.\ Astropart.\ Phys. {\bf 09}, 037 (2017).

\bibitem{ligo}
B.~P.~Abbott, et al. (LIGO Scientific Collaboration and Virgo Collaboration), Phys.\ Rev.\ Lett. {\bf 116}, 061102 (2016).

\bibitem{lesgourgues2}
A.~Cuoco, J.~Lesgourgues, G.~Mangano and S.~Pastor, Phys.\ Rev.\ D {\bf 71}, 123501 (2005).

\bibitem{neutribulk}
S.~R.~de~Groot, W.~A.~van~Leeuwen and C.~G.~van~Weert, Proc.\ K.\ Ned.\ Akad.\ Wet.\ B {\bf 82}, 113 (1979).

\bibitem{wcdm}
M.~Turner and M.~White, Phys.\ Rev.\ D {\bf 56}, 4439 (1997). 

\bibitem{lambda}
S.~Weinberg, Rev.\ Mod.\ Phys. {\bf 61}, 1 (1989); J.~Martin,  Comptes\ Rendus.\ Physique {\bf 13}, 566 (2012).

\bibitem{vcc}
Yin-Zhe~Ma, Nucl.\ Phys.\ B {\bf 804}, 262 (2008); V.~Sahni and A.~A.~Starobinsky, Int.\ J.\ Mod.\ Phys.\ D {\bf 9}, 373 (2000); A.~Strominger, Nucl.\ Phys.\ B {\bf 319}, 722 (1989); L.~R.~W.~Abramo, R.~H.~Brandenberger and V.~F.~Mukhanov, Phys.\ Rev.\ D {\bf 56}, 3248 (1997); I.~G.~Dymnikova and M.~Yu.~Khlopov, Mod.\ Phys.\ Lett.\ A {\bf 15}, 2305 (2000).

\bibitem{is}
W.~Israel, Ann.\ Phys. {\bf 100}, 310 (1976); W.~Israel and J.~M.~Stewart, Ann.\ Phys. {\bf 118}, 341 (1979).

\bibitem{nonlinear}
R.~Maartens and V.~M\'endez, Phys.\ Rev.\ D {\bf 55}, 1937 (1997).

\bibitem{velten}
H.~Velten, J.~Wang and X.~Meng, Phys.\ Rev.\ D {\bf 88}, 123504 (2013).

\bibitem{maartens}
R.~Maartens, arXiv:astro-ph/9609119.

\bibitem{noronha}
F.~S.~Bemfica, M.~M.~Disconzi and J.~Noronha, Phys.\ Rev.\ Lett. {\bf 122}, 221602 (2019). 

\bibitem{entropy}
J.~Garc\'\i a-Bellido and Ll.~Espinosa-Portal\'es, Phys.\ Dark\ Univ. {\bf 34}, 100892 (2021).

\bibitem{nucamendi}
F.~X.~Linares-Cede\~no and U.~Nucamendi, Phys.\ Dark\ Univ. {\bf 32}, 100807 (2021).

\bibitem{zimdahl}
W.~Zimdahl, J.~Triginer and D.~Pav\'on, Phys.\ Rev.\ D {\bf 54}, 6101 (1996).

\bibitem{equiv}
Ya.~B.~Zel'dovich, Sov.\ Phys.\ JETP\ Lett. {\bf 12}, 307 (1970); G.~L.~Murphy, Phys.\ Rev.\ D {\bf 8}, 4231 (1973); B.~L.~Hu, Phys.\ Lett.\ A {\bf 90}, 375 (1982).

\bibitem{zimdahl2}
W.~Zimdahl and D.~Pav\'on, Mon.\ Not.\ Roy.\ Astron.\ Soc. {\bf 266}, 872 (1994).

\bibitem{pavon}
B.~Wang, E.~Abdalla, F.~Atrio-Barandela and D.~Pav\'on, Rep.\ Prog.\ Phys. {\bf 87}, 036901 (2024); B.~Wang, E.~Abdalla, F.~Atrio-Barandela and D.~Pav\'on, Rep.\ Prog.\ Phys. {\bf 79}, 096901 (2016).

\bibitem{gamma}
M.~P.~Freaza, R.~S.~de~Souza and I.~Waga, Phys.\ Rev.\ D {\bf 66}, 103502 (2002); V.~H.~C\'ardenas, Eur.\ Phys.\ J.\ C {\bf 72}, 2149 (2012); R.~O.~Ramos, M.~Vargas~dos~Santos and I.~Waga, Phys.\ Rev.\ D {\bf 89}, 083524 (2014);  R.~C.~Nunes and D.~Pav\'on, Phys.\ Rev.\ D {\bf 91}, 063526 (2015);  V.~H.~C\'ardenas, M.~Cruz, S.~Lepe, S.~Nojiri and S.~D.~Odintsov, Phys.\ Rev.\ D {\bf 101}, 083530 (2020).

\bibitem{lr}
P.~H.~Frampton, K.~J.~Ludwick and R.~J.~Scherrer, Phys.\ Rev.\ D {\bf 84}, 063003 (2011).

\bibitem{lr2}
I.~Brevik, E.~Elizalde, S.~Nojiri and S.~D.~Odintsov, Phys.\ Rev.\ D {\bf 84}, 103508 (2011).

\bibitem{lr3}
I.~Albarran, M.~Bouhmadi-L\'opez, F.~Cabral and P.~Mart\'\i n-Moruno, J.\ Cosmol.\ Astropart.\ Phys. {\bf 11}, 044 (2015).

\bibitem{vagnozzi}
S.~Vagnozzi, S.~Dhawan, M.~Gerbino, K.~Freese, A.~Goobar and O.~Mena, Phys.\ Rev.\ D {\bf 98}, 083501 (2018).

\bibitem{moresco}
M.~Moresco~et~al., J.\ Cosmol.\ Astropart.\ Phys. {\bf 05}, 014 (2016).

\bibitem{eckart}
C.~Eckart, Phys.\ Rev. {\bf 58}, 267 (1940) ibid {\bf 58}, 919 (1940).

\bibitem{odintsov}
S.~Nojiri and S.~D.~Odintsov, Phys.\ Rev.\ D {\bf 72}, 023003 (2005); V.~H.~Cárdenas, M.~Cruz, S.~Lepe, S.~Nojiri and S.~D.~Odintsov, Phys.\ Rev.\ D {\bf 101}, 083530 (2020).

\bibitem{creation}
R.~Brout, F.~Englert and G.~Gunzig, Ann.\ Phys {\bf 115}, 78 (1978).

\bibitem{Suzuki:2012}
N.~Suzuki, et al. (The Supernova Cosmology Project), Astrophys.\ J. {\bf 746}, 85 (2012).

\bibitem{DES:2024}
M.~Vincenzi, et. al. (DES Collaboration), Astrophys.\  J. {\bf 975}, 86 (2024).

\bibitem{PantheonScolnic:2021}
D.~Scolnic, et. al. (The Pantheon+ Analysis: The Full Data Set), Astrophys.\  J. {\bf 938}, 113 (2022).

\bibitem{LIGOScientific:2017}
B.~P.~Abbott, et. al. (LIGO Collaboration), Nature  {\bf 551}, 7678 (2017).

\bibitem{Lopez:2022}
J.~Lopez, E.~Munive, A.~Avilez, O.~M.~Martinez, Mon.\ Not.\ Roy.\ Astron.\ Soc. {\bf 515}, 3199 (2022).

\bibitem{Alam:2017}
S.~Alam, et. al., Month.\ Not.\ Roy.\ Astron.\ Soc. {\bf 470}, 2617 (2017).

\bibitem{desi}
A.~G.~Adame, et al. (DESI Collaboration), arXiv:2404.03002 [astro-ph.CO].

\bibitem{shoes}
A.~G.~Riess, et al. (SH0ES Collaboration), Astrophys.\ J.\ Lett. {\bf 934}, L7 (2022).
\end{thebibliography}
\end{document}